\documentclass[twocolumn,showpacs,superscriptaddress,preprintnumbers,amsmath,amssymb,prc]{revtex4-2}

\usepackage{graphicx}
\usepackage{dcolumn}
\usepackage{bm}
\usepackage{float}
\usepackage{color}
\usepackage{ulem}
\usepackage{CJK}
\usepackage[table,usenames,dvipsnames]{xcolor}
\definecolor{linkcolor}{rgb}{0.6,0,0}
\definecolor{citecolor}{rgb}{0,0,0.75}
\definecolor{urlcolor}{rgb}{0.12,0.46,0.7}
\usepackage[breaklinks, colorlinks, urlcolor=urlcolor, linkcolor=linkcolor,citecolor=citecolor,pdfencoding=auto]{hyperref}
\hypersetup{linktocpage}
\usepackage{tabularx}

\usepackage{tikz,xcolor,hyperref}
\definecolor{lime}{HTML}{A6CE39}
\DeclareRobustCommand{\orcidicon}{
\begin{tikzpicture}
\draw[lime, fill=lime] (0,0) 
circle [radius=0.16] 
node[white] {{\fontfamily{qag}\selectfont \tiny ID}};
\draw[white, fill=white] (-0.0625,0.095) 
circle [radius=0.007];
\end{tikzpicture}
\hspace{-2mm}
}
\foreach \x in {A, ..., Z}{%
\expandafter\xdef\csname orcid\x\endcsname{\noexpand\href{https://orcid.org/\csname orcidauthor\x\endcsname}{\noexpand\orcidicon}}
}

\begin{document}
\begin{CJK*}{UTF8}{gbsn}

\title{Gravastar in the Framework of Symmetric Teleparallel Gravity}

\author{Sneha Pradhan\orcidA{}}
\email{snehapradhan2211@gmail.com}
    \affiliation{Department of Mathematics, Birla Institute of Technology and
Science-Pilani, \\ Hyderabad Campus, Hyderabad-500078, India}

\author{Sanjay Mandal\orcidB{}}
\email{sanjaymandal960@gmail.com}
\affiliation{Department of Mathematics, Birla Institute of Technology and
Science-Pilani, \\ Hyderabad Campus, Hyderabad-500078, India}

\author{P.K. Sahoo\orcidC{}}
\email{Corresponding author: pksahoo@hyderabad.bits-pilani.ac.in}
\affiliation{Department of Mathematics, Birla Institute of Technology and
Science-Pilani, \\ Hyderabad Campus, Hyderabad-500078, India}

\date{\today}

\begin{abstract}
In the current research, we present a novel gravastar model based on the Mazur-Mottola (2004) method with an isotropic matter distribution in $f(Q)$ gravity. The gravastar, a hypothesized substitute for a black hole, is built using the Mazur-Mottola mechanism. This approach allows us to define gravastar as having three stages. The first one is an inner region with negative pressure; the next region is a thin shell that is made up of ultrarelativistic stiff fluid, we have studied proper length, energy, entropy, and surface energy density for this region. Apart from that, we have demonstrated the possible stability of our suggested thin shell gravastar model through the graphical study of surface redshift. Exterior Schwarzschild geometry describes the outer region of the gravastar. In the context of $f(Q)$ gravity, we have discovered analytical solutions for the interior of gravastars that are free of any kind of singularity and the event horizon.
\\ \\
\textbf{Keywords:} Gravastar; modified gravity; non-metricity; Kuchowicz metric potential.

\end{abstract}

\maketitle

\section{Introduction}\label{sec:1}

The Schwarzschild solution is the most intriguing universal spherically symmetric static vacuum solution to the Einstein field equations under the background of general relativity (GR). The line element for  the Schwarzschild metric can be expressed as,

\begin{equation}
    ds^2= \left(1-\frac{2GM}{r}\right) dt^2-\frac{dr^2}{1-\frac{2GM}{r}}-r^2(d\theta^2+sin^2\theta d\phi^2),
    \label{eq:1}
\end{equation}
where $M$ is the gravitational mass of the object. A Black Hole (BH) of isolated mass $M$, with the static spherically symmetric line element \eqref{eq:1}, 
is known as Schwarzschild BH, where $R_S=2GM$ is known as Schwarzschild radius. BHs generates when the core of a massive star collapses toward the termination of its lifespan, are the most fascinating objects in modern astrophysics. In spite of being a successful theory, the Schwarzschild metric has these two significant flaws:
\begin{enumerate}
    \item The dynamical singularity at $r=0,\, 2\,GM$.
\item The presence of event horizon.
\end{enumerate}

To solve the aforementioned issues with BHs, Mazur and Mottola \cite{mazur2004gravitational,mottola2002gravitational}, 
 proposed a new solution that has no singularity, no event horizon, an ingenious idea of an extremely compact object, referred to as the Gravitationally Vacuum Condense Star or simply Gravastar. By employing the concept of Bose-Einstein condensation (BEC) to gravitational systems (i.e. a limited portion of the total number of particles starts to condense into the lowest-energy state below a certain temperature), they have put forth a novel theory to 
 explain the endpoint of gravitational collapse of a dying star's core. They create a cold, compact entity with an interior de Sitter condensate phase and an external Schwarzschild geometry of any total mass M,  which is removed from all the restrictions of the known classical black hole (CBH). As a result, this theory may be considered as a replacement concept for the CBH and has become a hot topic among researchers.
 
 According to Mazur and Mottola's model, the gravastar particularly contains three distinct regions with different equations of states (EoS).
 \begin{enumerate}
     \item An inner area with an isotropic de Sitter vacuum situation.
 \item  A thin shell of ultra-relativistic rigid fluid substance in between the interior and exterior region.
 \item The exterior region is entirely vacuum, and Schwarzschild geometry describes this situation correctly.
 \end{enumerate}
 
 Consequently, the following EoSs for different matters in these three regions can be used to define the complete gravastar system:

\begin{table}[H]
    \centering
    \begin{tabular}{|c|c|}
         \hline
  Region & EoS  \\
 \hline
 Interior ($0\leq r < r_1$)  & $p=-\rho$   \\
\hline
Shell ($r_1\leq r \leq r_2$) & $p=\rho$        \\
\hline
Exterior ($r_2<r$) &  $p=\rho=0$    \\
\hline
    \end{tabular}
    
\end{table}

There are two interfaces (junctions) at a distance $r_1$ and $r_2$ from the center. Where $r_1$ and $r_2$ denote the interior and exterior radius, respectively of the thin shell. To accomplish the necessary stability of the system assumed to be expanding by applying an inward force to balance the repulsion from within, the stiff matter must exist on the shell of thickness $r_1-r_2= \epsilon << 1$.

Although there are a few indirect pieces of evidence in the literature that can be utilized to indicate gravastar's existence and future confirmation, there is presently no observable evidence in favor of gravastar. Through the investigation of gravastar shadows, Sakai et al. \cite{sakai}  established a method to identify gravastar. As suggested by Kubo and Sakai \cite{kubo}, who asserted that black holes don't exhibit microlensing effects of maximal brightness, gravitational lensing is another method that might be used to find gravastars. The observation of GW150914 by interferometric LIGO detectors \cite{cardoso, Cardoso} suggested the existence of ringdown signals produced by sources without an event horizon.
A gravastar-like shadow has been seen in a recent examination of the picture taken by the First M87 Event Horizon Telescope (EHT) \cite{akiyama}.

There are several literary works on the gravastar that are based on various mathematical and scientific problems. However, the majority of these works are created within the context of general relativity. There has been a lot of work done on gravastar since Mazur and Mottola's idea. They demonstrated thermodynamic stability in their five-layer gravastar structure. The five-layer Mazur-Mottola model was simplified to three layers by Visser, and Wiltshire \cite{visser}, who also demonstrated its dynamical stability in the face of perturbations from spherically symmetric matter distributions or gravity fields. By examining relevant constraints for the stability of precise non-singular solutions of gravastars, Carter \cite{carter} expanded on this work.

To investigate the gravastar's interior Bili et al. \cite{bilic} replaced the de Sitter interior with a Chaplygin gas equation of state and regarded the system as a Born-Infeld phantom gravastar, whereas Lobo \cite{lobo} substitutes the interior vacuum with dark energy.
To overcome the singularity issue, Lobo and Arellano \cite{Lobo} linked the Schwarzschild outside with the internal nonlinear electrodynamic geometries. Anisotropic pressure in the "crust," which is essential for the development of structures like gravastars, was hypothesized by Cattoen et al. \cite{cattoen}. According to Ghosh et al. (\cite{ghosh}), it is impossible to extend a 4-dimensional gravastar to higher dimensions. Also, they have used the non-singular Kuchowicz metric potential to study several aspects of a gravastar \cite{Ghosh}.
Rahaman et al. \cite{rahaman} have built a gravastar model within the GR framework in (2 + 1) dimensions. However, Usmani et al. \cite{usmani} suggested a gravastar model with a charged interior that allowed the conformal motion and whose outer spacetime was defined by the Reissner-Nordström line element. Although Einstein's general relativity is one of the pillars of contemporary theoretical physics and has consistently been successful in revealing a great number of nature's hidden secrets, it has certain drawbacks from both a theoretical and an observational  standpoint. Theoretically, this idea has been challenged by observational evidence of the universe's accelerated expansion and the existence of dark matter. As a result, a number of different theories of gravity, including $f(R)$ gravity, $f(T)$ gravity, $f(R,T)$, $f(Q)$ gravity, etc. have been put out from time to time. All of these hypotheses can be regarded as essential for explaining the structure development and star system evolution in the universe.

Apart from the work on Einstein's GR, there are several works of gravastar in modified theories of gravity also.
Within the context of energy-momentum squared gravity, Sharif et al. \cite{sharif} explored the impact of the charge on the physical characteristics of gravastars and concluded that the non-singular solutions of the charged gravastar are physically feasible in  $f(R, T^2)$ gravity.
Das et al. \cite{das} have identified a collection of precise and singularity-free solutions to the gravastar that exhibits a number of physically significant as well as acceptable characteristics in $f(T)$ gravity. Sahoo and his group \cite{sahoo} investigated isotropic static spherically symmetric gravastars without charge in the Mazur-Motolla junction under the assumption of braneworld gravity.
In particular, the length of the shell, entropy, and energy have been studied as more realistic aspects of the gravastar model by Bhatti and his group \cite{bhatti} in $f(R,G)$ gravity. $f(R)$ gravity was first introduced in \cite{buchdahl}, and it is specifically represented by the arbitrary function of the Ricci scalar curvature $R$. However, there is another approach to explain gravitational interactions, referred to as torsion ($T$) and non-metricity ($Q$) is known as $f(T)$ and $f(Q)$ theories of gravity respectively. Weitzenböck and metric incompatible connections that are different from the GR Levi-Cevita connection are two examples of non-standard metric-affine connections that may be used to acquire both gravities. In this study, we focus on the modified symmetric teleparallel gravitational scenario (also known as $f (Q)$ gravity) to study the non-charged gravastar. There are lots of work on compact objects in the background of $f(Q)$ gravity. Mandal et al. \cite{mandal} studied the anisotropic compact stars in $f(Q)$ gravity with the quintessence field. In the work \cite{sokoliuk}, researchers have examined the physical behavior of strange stars using the MIT Bag EoS accepting Buchdahl metric for both linear and non-linear models.

Our research may be viewed as a singular study since we looked at a more generic isotropic model of uncharged gravastars and investigated the impact of non-metricity on both the interior and exterior solutions of the gravitating system in $f(Q)$ gravity. We have developed a set of advanced solutions for the three distinct areas with their respective EoS in the current study, which was inspired by the aforementioned publications. We analyzed surface redshift and the junction requirement for the creation of a thin shell to demonstrate the stability of our concept. We have looked into the physical features of gravastar while taking into consideration the three-layer model under Kuchowicz metric potential, unlike Visser and Wiltshire (\cite{visser}), who have simply reduced the Mazur, Mottola's five-layer construction into three layers to investigate the dynamical stability of gravastar.\\
In the current work, we looked at a few physical aspects of gravastars according to the $f(Q)$ theory of gravity and discovered analytical results for various phases of the structure. The paper is organized as follows:\\
In section-\ref{sec:1} we have given the introduction of gravastar model and recent work regarding that. In section-\ref{sec:2} we have discussed the mathematical formalism of $f(Q)$ gravity and have calculated the field equation in $f(Q)$ gravity for spherically symmetric metric. Three areas of the gravastar structure: the internal region, the thin shell, and the external region have been examined separately in section-\ref{sec:3}. In section-\ref{sec:4} we have studied  the physical features of thin shell to examine the stability of the model. The junction condition has been investigated (along with EoS) in section-\ref{sec:5}. Finally, we have concluded regarding the stability of our model in section-\ref{sec:6}.

\section{Basic mathematical formalism of $f(Q)$ gravity}\label{sec:2}

For the purposes of STEGR, we assume that the gravastar under consideration exists on the differentiable Lorentzian manifold $\mathcal{M}$, which may be adequately characterized by the metric tensor $g_{\mu\nu}$, its determinant $g$, and its affine connection $\Gamma$:
\begin{equation}
    g=g_{\mu\nu}dx^\mu\otimes dx^\nu,
\end{equation}

Here  $\Gamma^\alpha_{\,\,\,\,\beta}$ is the connection one form, which may be represented in terms of one forms of contortion tensor, disformation, and connection of Levi-Civita
\cite{Ortin}:

\begin{equation}
\Gamma^{\alpha}_{\,\,\,\,\beta}=w^{\alpha}_{\,\,\,\,\beta}+K^{\alpha}_{\,\,\,\,\beta}+L^{\alpha}_{\,\,\,\,\beta}.
\end{equation}

The aforementioned expression can be represented as follows:
\begin{equation}
\Gamma^{\alpha}_{\,\,\,\,\mu\nu}=\gamma^{\alpha}_{\,\,\,\,\mu\nu}+K^{\alpha}_{\,\,\,\,\mu\nu}+L^{\alpha}_{\,\,\,\,\mu\nu},
    \label{eq:2}
\end{equation}
where the variables $\gamma$, $K$, and $L$ in the equation \eqref{eq:2} are the affine connection that is consistent with the Levi-Cevita metric, contortion, and disformation tensors respectively. Assuming $f(Q)$ gravity formalism, the symmetric teleparallelism that results from the non-metricity one form and associated tensor completely describes the gravitational sector :
\begin{equation}
    Q^\alpha_{\,\,\,\,\beta}=\Gamma_{(ab)},\quad Q_{\alpha\mu\nu}=\nabla_\alpha g_{\mu\nu},
\end{equation}
where the symmetric portion of the tensor has the following definition :
\begin{equation}
    F_{(\mu\nu)}=\frac{1}{2}\bigg(F_{\mu\nu}+F_{\nu\mu}\bigg).
\end{equation}
If contortion disappears, the only thing remaining will be disformation tensor terms as:
\begin{equation}
    Q_{\alpha\mu\nu}=-L^{\beta}_{\,\,\,\,\alpha\mu}g_{\beta\nu}-L^{\beta}_{\,\,\,\,\alpha\nu}g_{\beta\mu},
\end{equation}
The following formula defines the  disformation tensor:
\begin{equation}
    L^{\alpha}_{\mu\nu}=\frac{1}{2}Q^{\alpha}_{\mu\nu}-Q_{(\mu\nu)}^{\alpha}.
    \label{eq:3}
\end{equation}
The non-metricity scalar, which is made up of non-metricity and the so-called superpotential, is the primary gravitational quantity in the STEGR formalism.
\begin{equation}
    Q=-P^{\alpha\mu\nu}Q_{\alpha\mu\nu},
\end{equation}
where the following complicated form describes the superpotential:
\begin{equation}
    P^{\alpha}_{\,\,\,\,\mu\nu}=\frac{1}{4}\bigg[2Q^{\alpha}_{\,\,\,\,(\mu\nu)}-Q^{\alpha}_{\,\,\,\,\mu\nu}+Q^\alpha g_{\mu\nu}-\delta^{\alpha}_{(i}Q_{j)}-\overline{Q}^\alpha g_{\mu\nu}\bigg].
\end{equation}
Where, $Q^{\alpha}=Q^\nu_{\,\,\,\,\alpha\nu}$ and $\overline{Q}_\alpha=Q^\mu_{\,\,\,\,\alpha\mu}$ are two independent traces of the non-metricity tensor $Q_{\alpha\mu\nu}=\nabla_\alpha g_{\mu\nu}$. Since we have covered all the terms required by the STEGR formalism, we can go ahead to provide the modified action integral for $f(Q)$ gravity \cite{Xu}:
\begin{equation}
    \mathcal{S}[g,\Gamma,\Psi_i]=\int d^4x \sqrt{-g}f(Q)+\mathcal{S}_{\mathrm{M}}[g,\Gamma,\Psi_i].
    \label{eq:3}
\end{equation}
In this case, $\mathcal{S}_{\mathrm{M}}[g,\Gamma,\Psi_i]$ represents the action integral which signifies the contribution of an extra matter field $\Psi_i$ that is both minimally and maximally linked to gravity to the overall Einstein-Hilbert action integral. The abstract field equations for the theory may be obtained by varying the equation \eqref{eq:3} with regard to the metric tensor inverse $g^{\mu\nu}$ :
\begin{equation}
\begin{gathered}
\frac{2}{\sqrt{-g}}\nabla_\gamma\left(\sqrt{-g}\,f_Q\,P^\gamma\;_{\mu\nu}\right)+\frac{1}{2}g_{\mu\nu}f \\
+   f_Q\left(P_{\mu\gamma i}\,Q_\nu\;^{\gamma i}-2\,Q_{\gamma i \mu}\,P^{\gamma i}\;_\nu\right)=-T_{\mu\nu},
\label{eq:12}
\end{gathered}
\end{equation}
here $f_Q\equiv\frac{df}{dQ}$ and the widely used energy momentum tensor $T_{\mu\nu} $, whose generic form might be represented as:
\begin{equation}
    T_{\mu\nu}=-\frac{2}{\sqrt{-g}}\frac{\delta(\sqrt{-g} \mathcal{L}_{\mathrm{M}})}{\delta g^{\mu\nu}}.
\end{equation}
Where The Lagrangian density of matter fields is represented by the symbol $\mathcal{L}_{\mathrm{M}}$ satisfying the relation $\int d^4x \sqrt{-g} \mathcal{L}_{\mathrm{M}}=\mathcal{S}_{\mathrm{M}}[g,\Gamma,\Psi_i]$. Additionally, by changing the action in relation to the affine connection $\Gamma^\alpha_{\,\,\,\,\mu\nu}$, we get :
\begin{equation}
\nabla_\mu \nabla_\nu \left(\sqrt{-g}\,f_Q\,P^\gamma\;_{\mu\nu}\right)=0.
\label{eq:14}
\end{equation}
In the next subsection,  for spherically symmetric objects we will derive the precise form of field equations (\ref{eq:12}) and (\ref{eq:14}) written above.

\subsection{Motion equations for spherically symmetric objects in $f(Q)$ gravity}

Within $f(Q)$ gravity the field equation or motion equation for a spherically symmetric object takes the following form:\\
 \begin{equation}
 \begin{gathered}
     \kappa T_{tt} = \frac{e^{\nu-\lambda}}{2r^2}  [2rf_{QQ} Q'(e^{\lambda} -1)+ f_{Q}[(e^{\lambda} -1)\\(2+r\nu')+(1+e^{\lambda})r \lambda']+fr^{2}e^{\lambda}] ,
 \end{gathered}
 \end{equation}
  
   \begin{equation}
   \begin{gathered}
       \kappa T_{rr} =- \frac{1}{2r^2}  [2rf_{QQ} Q'(e^{\lambda} -1)+ f_{Q}[(e^{\lambda} -1)\\(2+r\nu'+r\lambda')-2r\nu']+fr^{2}e^{\lambda}], 
   \end{gathered}
 \end{equation}
 
  \begin{equation}
  \begin{gathered}
      \kappa T_{\theta\theta} = - \frac{r}{4e^{\lambda}}  [-2rf_{QQ} Q' \nu' + f_{Q}[2\nu'(e^{\lambda} -2)\\ -r\nu'^{2}+\lambda'(2e^{\lambda}+r\nu')-2r\nu'']+2fre^{\lambda}], \\
  \end{gathered}
  \end{equation}
where the line element for the spherically symmetric metric is defined as :\\
 \begin{equation}
    ds^2= e^{\nu} dt^2-e^{\lambda} dr^2-r^2(d\theta^2+ sin^2 \theta d\phi^2).
    \end{equation}
As a result, the usual energy-momentum tensor for perfect fluid matter distribution can be written as follows:
\begin{equation}
        T_{\mu\nu}=diag(e^{\nu}\rho, e^{\lambda}p,r^{2} p, r^2 p sin^{2}\theta).
    \end{equation}
  Applying the values of the energy-momentum tensor components to the aforesaid field equations now yields the following result: \\
  \begin{equation}
  \begin{gathered}
    8\pi\rho = \frac{1}{2r^2 e^{\lambda}}  [2rf_{Q Q} Q'(e^{\lambda} -1)+ f_{Q}[(e^{\lambda} -1)(2+r\nu')\\+(1+e^{\lambda})r \lambda']+fr^{2}e^{\lambda}] ,
     \label{eq:18}
  \end{gathered}
  \end{equation}

 \begin{equation}
    \begin{gathered}
    8\pi p = - \frac{1}{2r^2 e^{\lambda}}  [2rf_{Q Q} Q'(e^{\lambda} -1)+ f_{Q}[(e^{\lambda} -1)\\(2+r\nu'+r\lambda')-2r\nu']+fr^{2}e^{\lambda}] ,
     \end{gathered}
\label{eq:19} 
 \end{equation}

\begin{equation}
    \begin{gathered}
      8\pi p = - \frac{1}{4re^{\lambda}}  [-2r f_{Q Q} Q' \nu' + f_{Q}[2\nu'(e^{\lambda} -2)- \\ r\nu'^{2}+\lambda'(2e^{\lambda}+r\nu')-2r\nu'']+2f re^{\lambda}].
      \end{gathered}
\label{eq:20}
\end{equation}
Therefore, the mathematical form of non-metricity scalar is,
\begin{equation}
    Q =\frac{1}{r}(\nu'+\lambda')(e^{-\lambda}-1).
\end{equation}

\section{Gravastar Condition}\label{sec:3}

By carefully determining the metric potential $e^{\lambda}$ with the use of the field equations \eqref{eq:18}-\eqref{eq:20}, we have explored three distinct regions of the gravastar. In order to determine the metric potential $e^{\lambda}$ we have taken the physically feasible, singularity-free metric potential, specifically the Kuchowicz metric potential of form \cite{kuchowicz},
\begin{equation}
    e^{\nu}=e^{{B r^2}+ 2 ln C}
    \label{eq:24}
\end{equation}
 to evaluate such a compact spherically symmetric astronomical object like a gravastar accurately.
Here $B$ and $C$ are arbitrary constants. $C$ is a dimensionless parameter and $B$ is of dimension [$L^{-2}$]. Using the aforementioned condition\eqref{eq:24} we have simplified the field equation \eqref{eq:18}-\eqref{eq:20} for linear model $f(Q)=a Q + b$, where $a$ and $b$ are the model parameter. The simplified forms are,

\begin{equation}
    \rho=\frac{e^{-\lambda }   \left(-2 a+e^{\lambda } \left(2 a+b r^2\right)+2 a r \lambda '\right)}{16 \pi r^2}
    \label{eq:25},
\end{equation}
\begin{equation}
  p=  \frac{1}{16 \pi}   \left(-b-\frac{2 a e^{-\lambda } \left(-1+e^{\lambda }-2 B r^2\right)}{r^2}\right)
    \label{eq:26},
\end{equation}

\begin{gather}
   p=-\frac{  e^{-\lambda (r)} \left(a \left ( B r^2+1\right) \lambda '(r)-2 a B r \left(B r^2+2\right)+b r e^{\lambda (r)}\right)}{16 \pi r}
    \label{eq:27}.
\end{gather}

Finally, in $f(Q)$ gravity the energy conservation equation for a line element in (3+1) dimension can be written as,
\begin{equation}
    \frac{dp}{dr}+ \frac{\nu'}{2}(p+\rho)=0
    \label{eq:28}.
\end{equation}
It is clear from the aforementioned equation that the gravitational force must balance out the hydrostatic force, or pressure gradient, for a gravitating system to be in equilibrium.

\subsection{Interior region}

 We presumed the  EoS for the interior area as stated in Mazur, Mottola's work in \cite{mazur2004gravitational,mottola2002gravitational},
\begin{equation}
    p=-\rho.
    \label{eq:29}
\end{equation}
 This EoS is referred to as the dark energy EoS \cite{Perlmutter}-\cite{Ray} with the parametric value $\omega=-1$.
The inward gravitational attraction of the shell is balanced by this negative (repulsive) pressure acting radially outwards from the center of the spherically symmetric gravitating system. From the equations \eqref{eq:28},\eqref{eq:29} it can be written that\\
\begin{equation}
    p=-\rho=-\rho_c,
    \label{eq:30}
\end{equation}
where $\rho_c$ is the critical density of the gravastar.
With the help of \eqref{eq:30} we get the metric potential $e^{-\lambda}$ from equation \eqref{eq:25} as:
\begin{equation}
    e^{-\lambda}=1+\frac{r^2}{6a }( b-16 \pi \rho_c)-\frac{c_1}{2ar}.
\end{equation}
To get the regular solution at the center we could take $c_1=0$. So,
\begin{equation}
    e^{-\lambda}=1+\frac{r^2}{6a }( b-16 \pi \rho_c).
    \label{eq:33}
\end{equation}
For making our solution free from central singularity we have taken $\frac{r^2}{6a }( b-16 \pi \rho_c)\neq -1$. Thus we get two non-singular space-time metrics for describing the internal structure of the gravastar system.
Also, the following equation can be used to determine the active gravitational mass of the internal area,
\begin{equation}
    M(R)=\int_{0}^{R_1=R} 4 \pi r^2 \rho dr= \frac{4\pi R^3 \rho_c}{3}
\end{equation}
where $R_1$ is the boundary of the interior region which is set as $R$, and r is the radial coordinate.

\begin{figure}[h]

\includegraphics[scale=0.45]{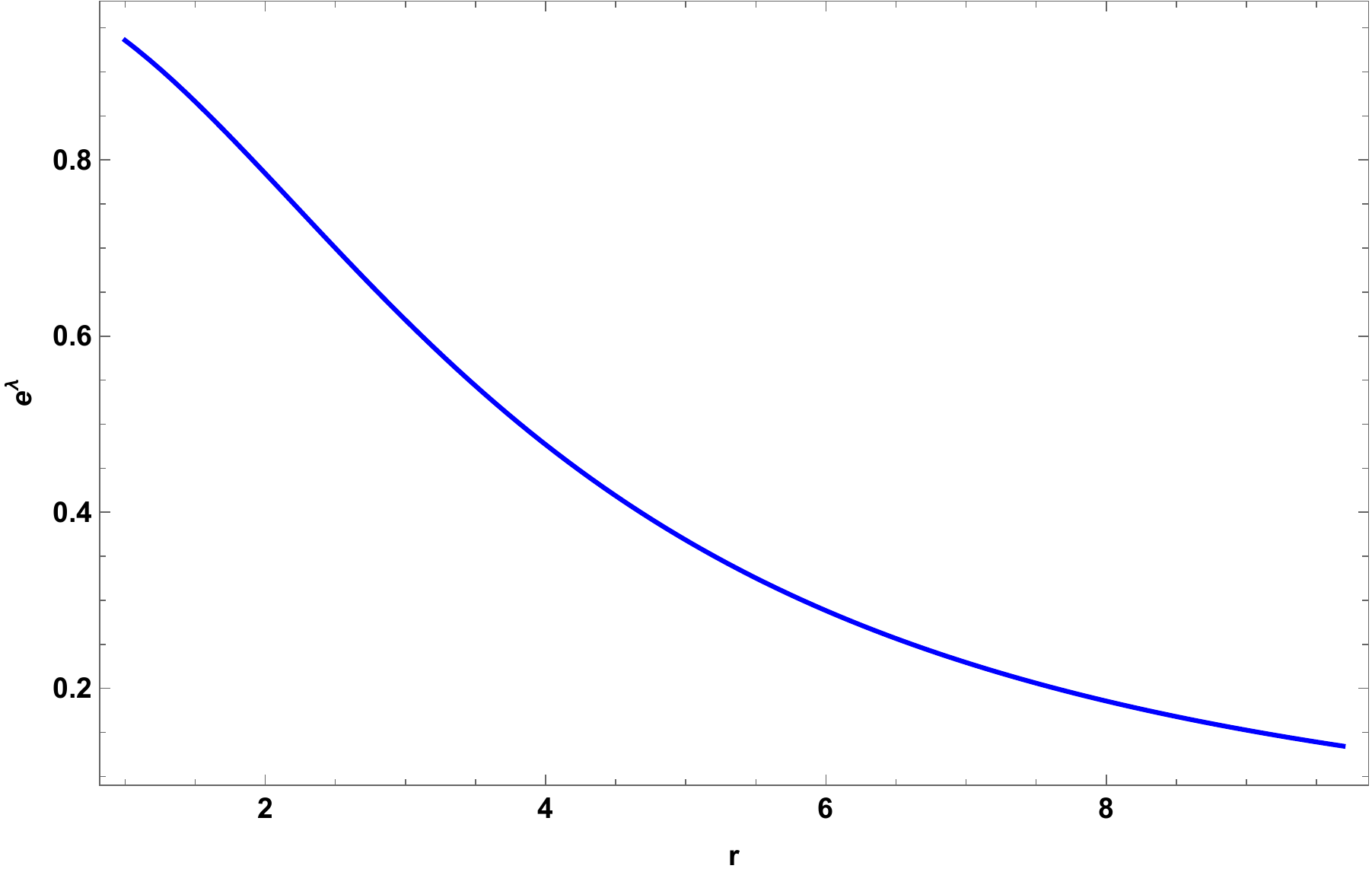}
    \caption{ Graphical analysis of the metric potential within the interior region with the radial coordinate r(km) for $a=1.7$, $b=0.7$.}
    \label{fig:1}
\end{figure}
 The evolution of space-time metric $e^{\lambda}$ with the radial parameter $r$ has been shown in Fig.\ref{fig:1}. From the figure, we can physically interpret that the metric potential remains positive throughout the region, and it is regular at $r=0$, also finite, and has no central singularity. 

\subsection{Intermediate thin shell}

The shell is made of ultrarelativistic fluid, and it abides by the EoS \begin{equation}
p=\rho    .
\end{equation}
Zel'dovich \cite{Zeldovich} was the one who originally came up with the concept of this form of fluid  in relation to the cold baryonic cosmos, also known as the stiff fluid.
By equating the isotropic pressure and matter density of the gravastar we get the space-time metric $e^{\lambda}$ for the shell region as,

\begin{equation}
    e^{\lambda}=\frac{2a B^2 e^{Br^2}r^2}{e^{Br^2}(2aB+b(-1+Br^2))-8B^2(b-2aB)C_1}.
    \label{eq:36}
\end{equation}
The boundary condition could be used to determine the value of integrating constants $C_1$. 

\begin{figure}[h]
\includegraphics[scale=0.456]{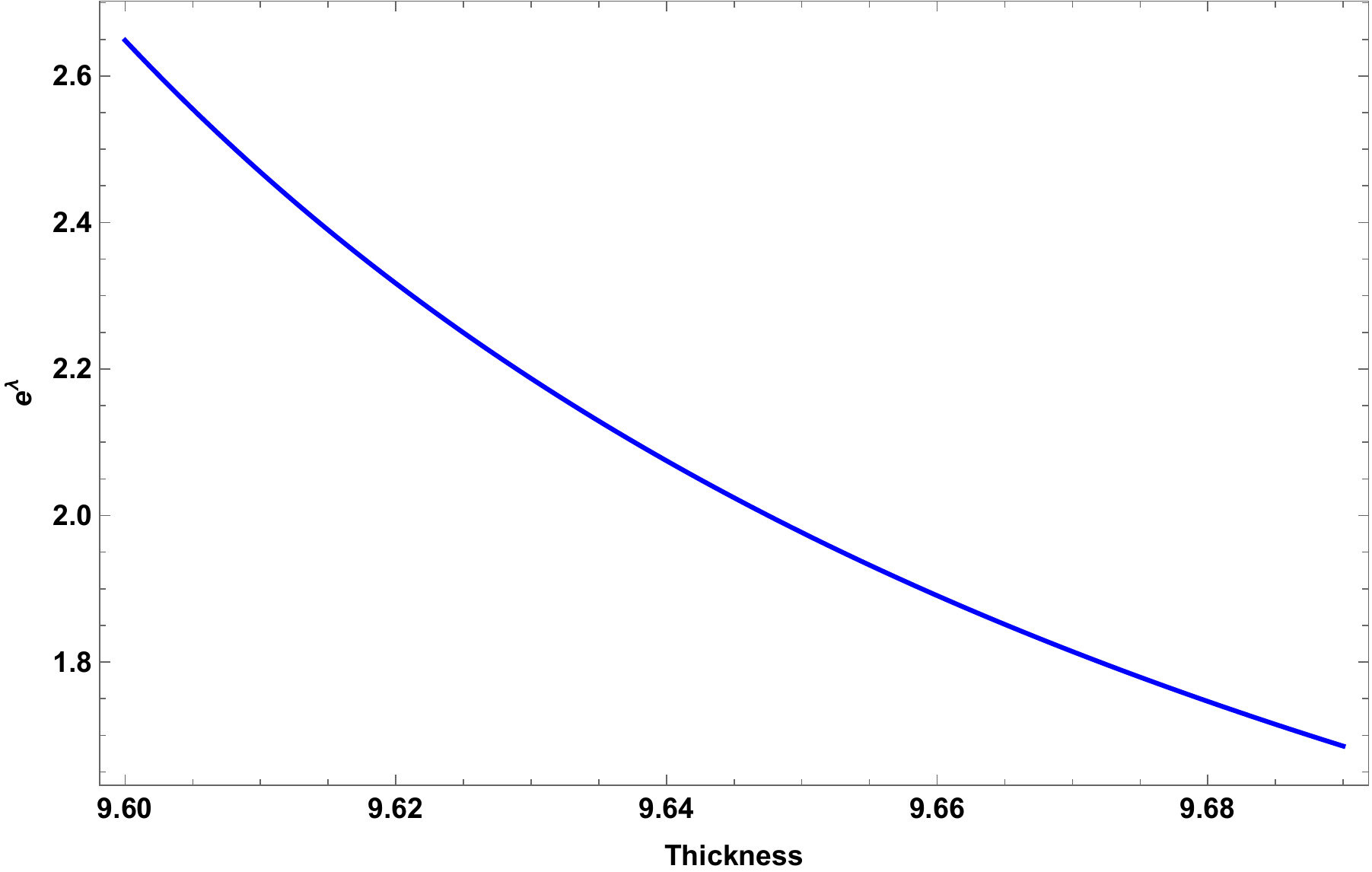}
    \caption{Graphical analysis of the metric potential within the shell with its thickness($\epsilon$ in km) for $a=1.7$ and $b=0.7$.}
    \label{fig:2}
\end{figure}

The variation of metric potential $e^{\lambda}$ for the shell region has shown in Fig. \ref{fig:2}. From the figure, we can say that the space-time metric within the shell region shows the same behavior as the interior region, which establishes a positive behavior in our solution.

\subsection{Exterior region}

It is believed that the gravastar's exterior is fully vacuum outside of the shell and follows the EoS  $p=\rho=0 $ rule. Therefore the Schwarzschild metric,
\begin{equation}
    ds^2= \left(1-\frac{2M}{r}\right) dt^2-\frac{dr^2}{1-\frac{2M}{r}}dr^2-r^2(d\theta^2+sin^2\theta d\phi^2)
\end{equation}
describes the outside geometry of the gravastar correctly. Now for vacuum EoS, we have determined the metric potential $e^{\lambda}$ as,

\begin{equation}
    e^{\lambda}=\frac{2(a+2aBr^2)}{2a+br^2}
    \label{eq:38}.
\end{equation}
Finally, we have three space-time  metrics \eqref{eq:33},\eqref{eq:36},\eqref{eq:38} for three different regions.
\subsubsection{Boundary Condition:}
In order to determine the value of three constants $B,\, C,\, C_1$ we will equate our determined  metric potential $e^{\lambda}$ at the boundaries. There are two limits in the gravastar configuration: one is between the interior and intermediate thin shell at a distance $R_1$, and the other is between the shell and outer space at a distance $R_2$ from the center. The metric functions at these interfaces must be continuous for any stable system. Therefore we have matched these metric potentials with outer space-time. Finally, we obtain the following form for these constants.

\begin{equation}
     B=\frac{4aM+bR^{3}_2}{4aR_2^{2}(R_2-2M)},
\end{equation}
\begin{equation}
     C_1=\frac{e^{BR_2^2}(b-2aB-4aB^2MR_2-bBR_2^2+2aB^2R_2^2)}{8B^2(2aB-b)},
\end{equation}
  \begin{equation}
      C^2=e^{-BR_2^2}(1-\frac{2M}{R_2}).
  \end{equation}

Here we have taken PSR J1416-2230 compact star with the total mass $M=1.97 M_{\odot}$, and outer radius $R_2=9.69$ km. Thus the ratio $\frac{2M}{R}<1$ holds for gravastar configuration. Also, we have taken the value of critical density $\rho_c=0.0001$ and $\rho_0=1$. Putting the values of the above quantity along with the model parameter $a=1.7$ and $b=0.7$ we have obtained the numerical values of the constants as, $B=0.17301$ $\text{km}^{-2}$, $C_1=7.3465\times10^9$, $C^2=1.80065\times10^{-8}$. Now the question is why we choose the model parameter like this. The reason is the nature of the constant $B$ and $C^2$. Here the constant $B$ is of dimension $l^{-2}$, so it should take a positive value. Also, the term $C^2$ is a positive quantity. Therefore for making the positive numerical values of these constants, we have taken such a specific value of the model parameter. We have verified that for the aforementioned condition, the model shows similar behavior for different values of $a$ and $b$.  In order to determine the feasibility of our gravastar model, we will verify it through some physical features of the shell region in the next section.

\section{Some physical features of the shell region}\label{sec:4}

\subsubsection{Pressure and matter density}
One of the leading indicators of a spherically symmetric object's physical qualities is the variation of density with radial coordinates.
For the shell region using the EoS $p=\rho$ from the equations \eqref{eq:28} one can easily get,

\begin{equation}
    p=\rho=\rho_0 e^{-B r^2}.
\end{equation}

\begin{figure}[h]
\advance\leftskip-1.1cm
\includegraphics[scale=0.45]{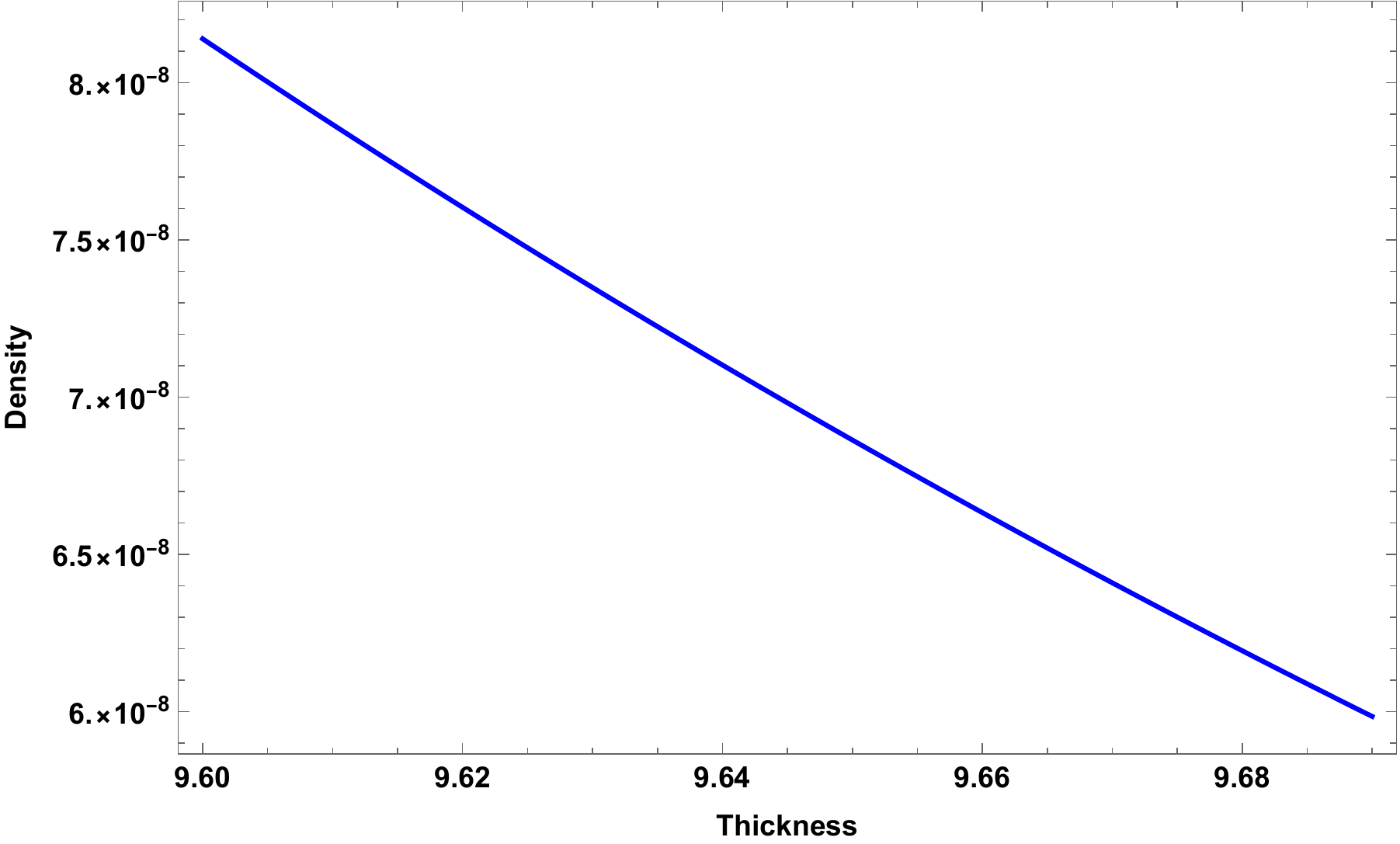}
    \caption{ Graphical analysis of the pressure or matter density of the shell with its thickness($\epsilon$ in km) for $a=1.7$ and $b=0.7$.}
    \label{fig:3}
\end{figure}

The evolution of density with respect to the thickness of the shell has shown in Fig.(\ref{fig:3}). One can see from Fig.(\ref{fig:3}) that both pressure and matter density stay positive throughout the shell's range and rapidly decreases with the increases in thickness.

\subsubsection{Proper length}
 In accordance with the hypothesis put forward by Mazur and Mottola \cite{mazur2004gravitational,mottola2002gravitational}, the phase boundary of the interior area is defined by the stiff fluid shell  at $r=R$, and its proper thickness $\epsilon$ may be considered to be extremely small i.e. $\epsilon<<1$. Consequently, the lower limit of the outer region has been taken as $r=R+\epsilon$. The proper thickness / proper length between two interfaces, or that of the shell, may be written as,

    \begin{gather}
     l=\int^{R+\epsilon}_{R} \sqrt{e^{\lambda}}dr 
 \\
     =\int^{R+\epsilon}_{R} \sqrt{\frac{2a B^2 e^{Br^2}r^2}{e^{Br^2}(2aB+b(-1+Br^2))-8B^2(b-2aB)C_1}}dr.
     \end{gather}

     The above expression is  very difficult to integrate analytically. Let us assume $f(r)$ as the primitive of $\sqrt{e^{\lambda}}$. Then by the fundamental theorem of calculus, we have,
\begin{gather}
   l=\int^{R+\epsilon}_{R} \frac{d}{dr}f(r)dr=f(R+\epsilon)-f(R) \\
    =\left(f(R)+\epsilon f'(R) +\frac{\epsilon^2}{2!} f''(R)+.....\right)-f(R)\\=\epsilon f'(R) +\frac{\epsilon^2}{2!} f''(R)+......
\end{gather}
    
We have used Taylor series expansion to calculate the thickness of the thin shell and confined our calculations to the first-order term of the thickness parameter $\epsilon$. Finally, we have determined the precise thickness of the thin shell as,

\begin{equation}
    l=\epsilon\sqrt{\frac{2a B^2 e^{BR^2}R^2}{e^{BR^2}(2aB+b(-1+BR^2))-8B^2(b-2aB)C_1}}.
\end{equation}

\begin{figure}[h]
\advance\leftskip-1.3cm
\includegraphics[scale=0.456]{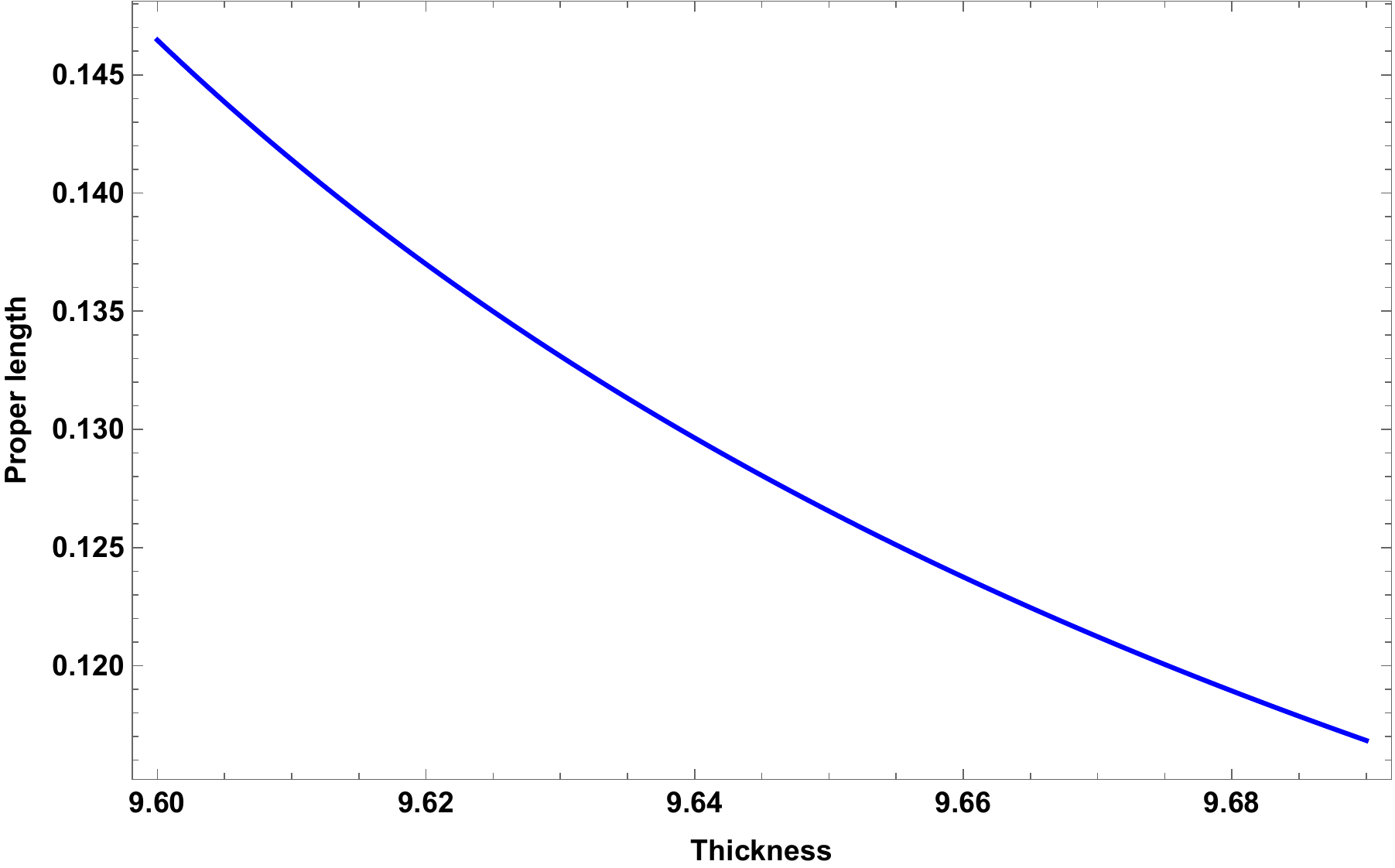}
    \caption{ Graphical analysis of the proper length of the shell with its thickness($\epsilon$ in km) for $a=1.7$ and $b=0.7$.}
    \label{fig:4}
\end{figure}

We have plotted numerically the proper length variations with the thickness parameter $\epsilon$ in Fig \ref{fig:4}. It is observed that when thickness values increase, the proper length of the shell steadily decreases.

\subsubsection{Energy:}
One may argue that this implies the negative energy area in the inner region when we consider the EoS, $p=-\rho$, proving the core's repulsive character. However, it can be determined that the energy contained in the shell is,
\begin{gather}
    E=\int^{R+\epsilon}_{R} 4 \pi r^2 \rho dr=\int^{R+\epsilon}_{R}4\pi r^2 \rho_0 e^{-Br^2} dr\\=4 \pi  \left[-\frac{e^{-B r^2} r}{2 B}+\frac{\sqrt{\pi } \text{Erf}\left[\sqrt{B} r\right]}{4 B^{3/2}}\right]^{R+\epsilon}_R.
\end{gather}

\begin{figure}[ ]
\advance\leftskip-1.3cm
\includegraphics[scale=0.456]{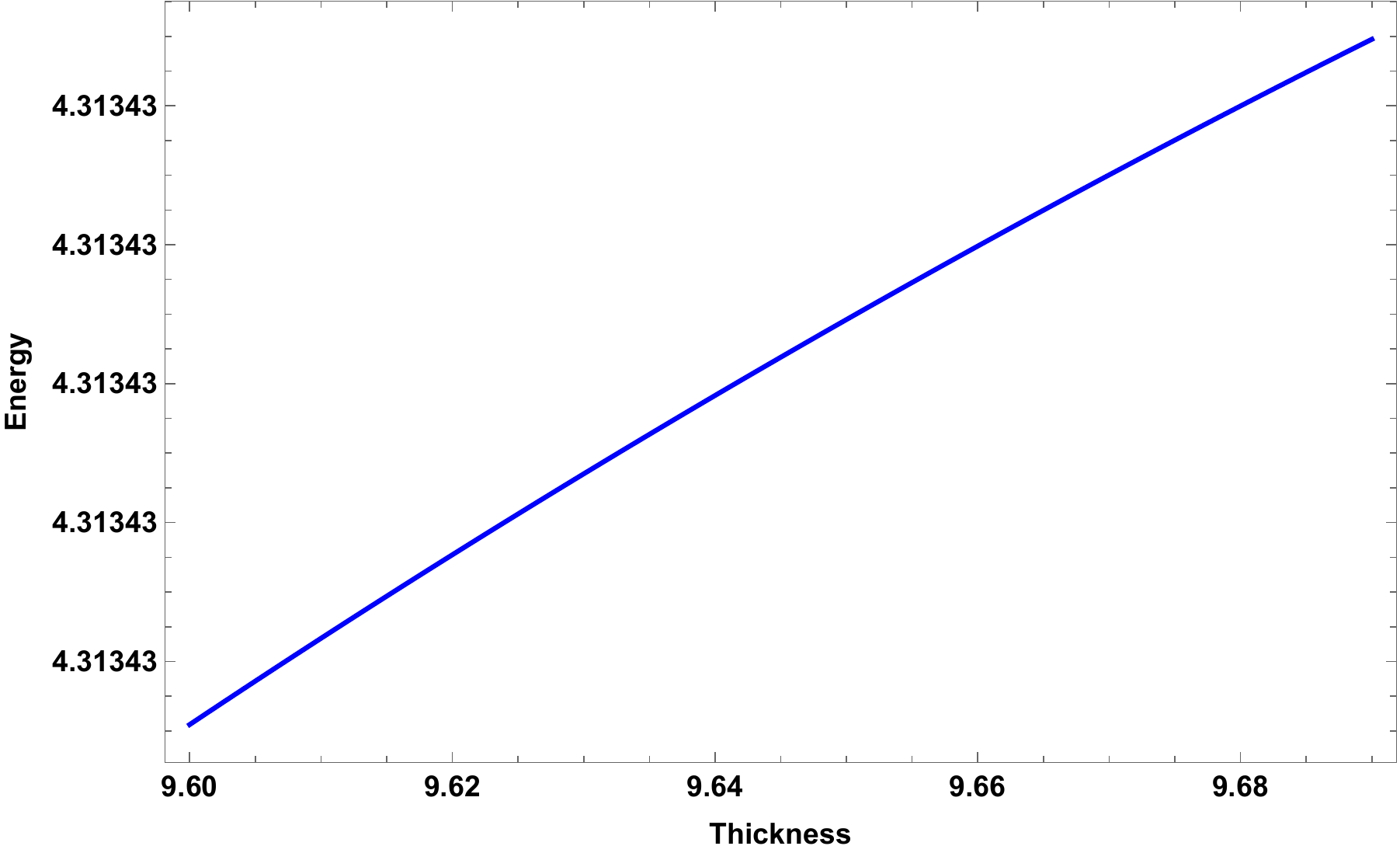}
    \caption{ Graphical analysis of the surface energy of the shell with its thickness($\epsilon$ in km) for $a=1.7$ and $b=0.7$.}
    \label{fig:5}
\end{figure}
From Fig. \ref{fig:5} it is discovered that the energy variation for the shell is positive and monotonically increases toward the outer surface. This implies that the shell's outer boundary is denser than its internal limit. It meets with the condition that the shell's energy grows as the radial distance increases.

\subsubsection{Entropy}
The interior of the gravastar must have zero entropy density, which is stable for the single condensate area, according to Mazur and Mottola \cite{mazur2004gravitational, mottola2002gravitational}. Entropy on the shell, however, is typically not zero. The following formula makes it simple to compute the entropy of the relativistic star system (static) gravastar:

\begin{equation}
   S= \int^{R+\epsilon}_{R} 4 \pi r^2 s(r) \sqrt{e^{\lambda}} dr,
\end{equation}
where $s(r)=\alpha \sqrt{p/{2\pi}}$ is the entropy density and $\alpha$ is the dimensionless parameter. Applying the same technique once again using the Taylor series approximation up to first order term of $\epsilon$, such as determining the proper length, we have determined the entropy of the shell as,\\
\begin{multline}
     S=4\epsilon  \alpha \pi  R^2 \times \\
     \sqrt{\frac{2a B^2 R^2}{e^{BR^2}(2aB+b(-1+BR^2))-8B^2(b-2aB)C_1}}
\end{multline}

\begin{figure}[h]
\advance\leftskip-1.3cm
\includegraphics[scale=0.456]{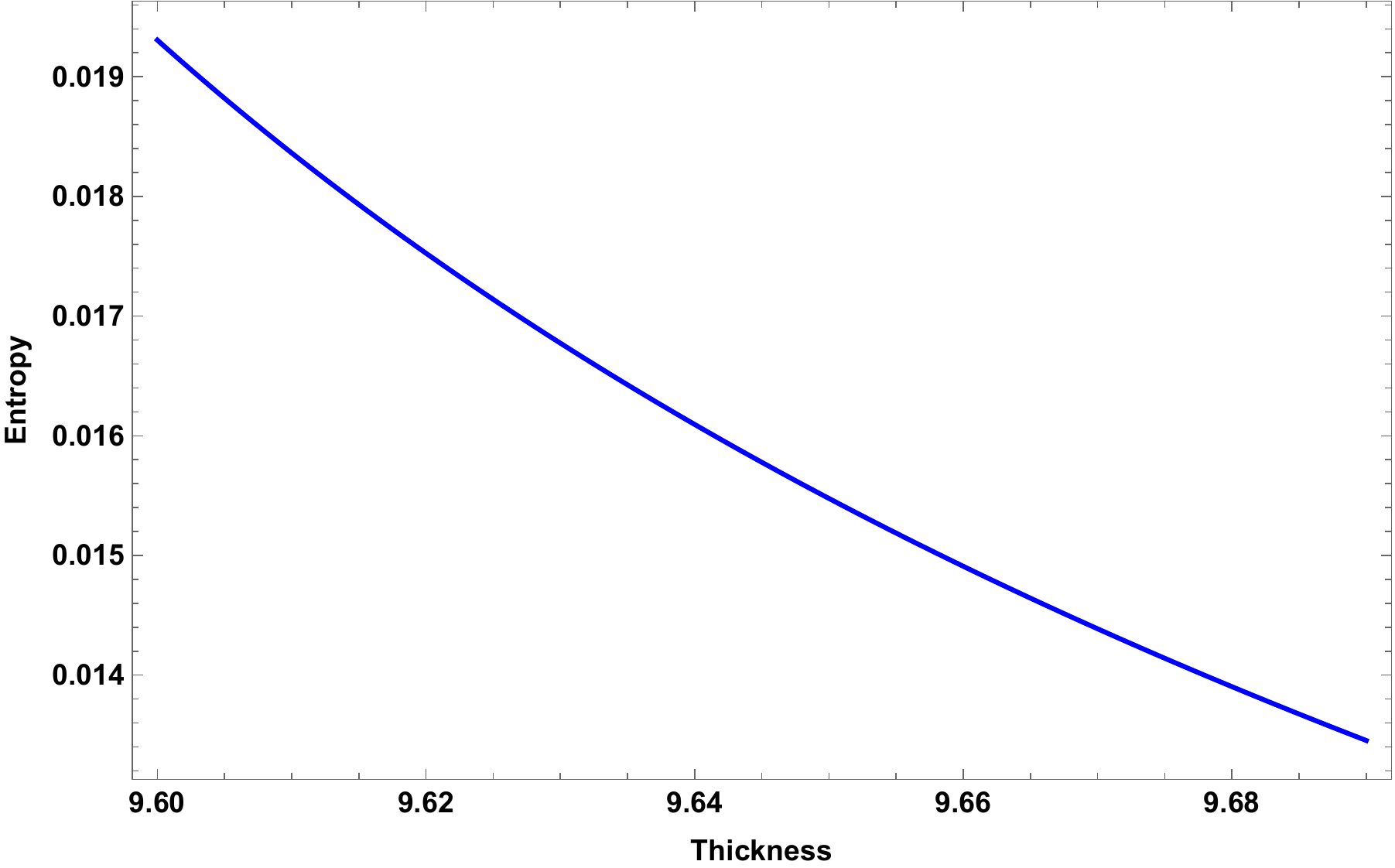}
    \caption{ Graphical analysis of the entropy of the shell with its thickness($\epsilon$ in km) for $a=1.7$ and $b=0.7$.}
    \label{fig:6}
\end{figure}
The variation of the shell entropy has shown in Fig. \ref{fig:6}. It is noted that the entropy of the shell decreases as its thickness increases in $f(Q)$ gravity. This behavior, however, does not suggest that our solution is unstable.

\subsubsection{Surface Redshift}
\begin{figure}[h]
\advance\leftskip-1.3cm
\includegraphics[scale=0.456]{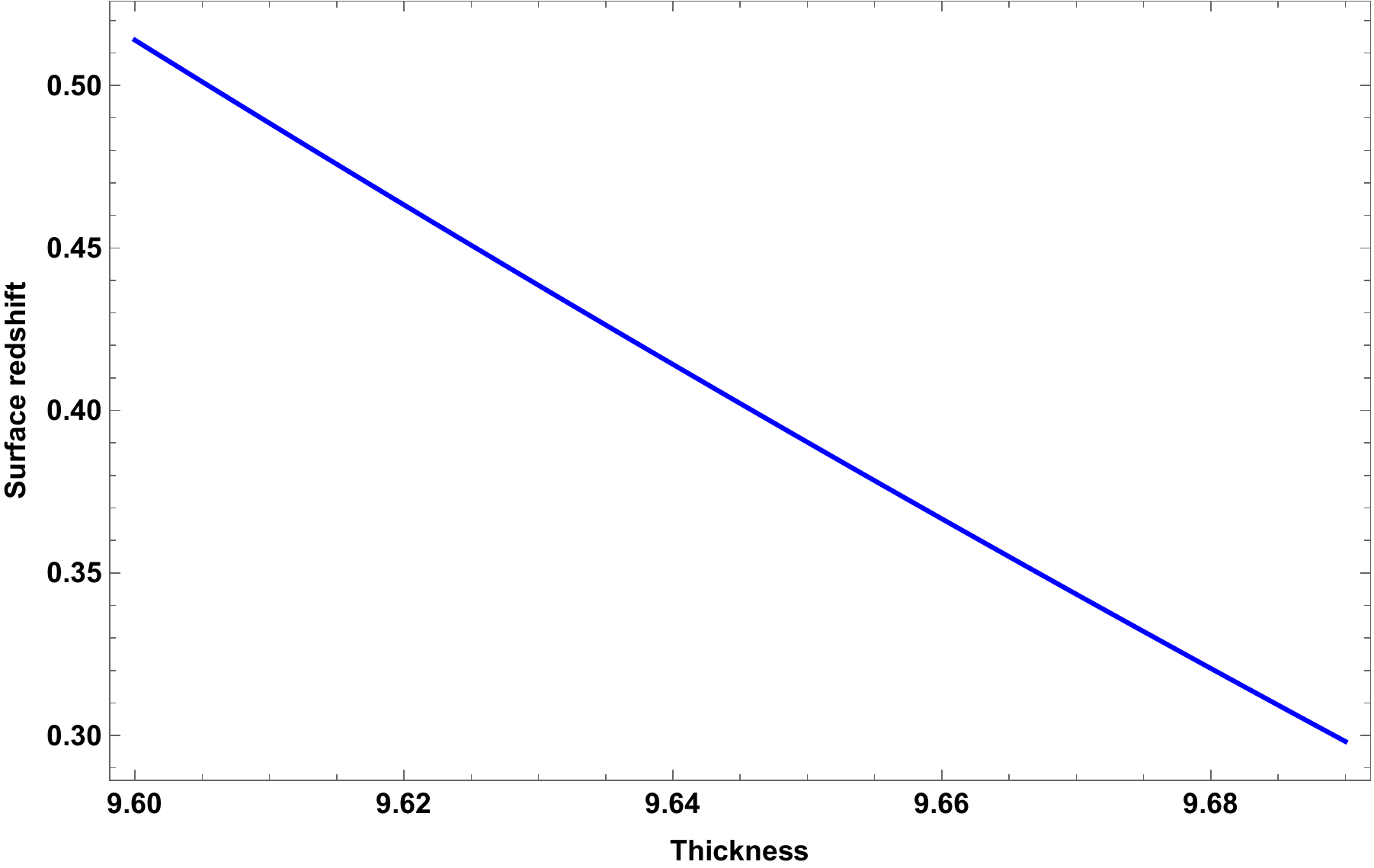}
    \caption{ Graphical analysis of the surface redshift of the shell with its thickness($\epsilon$ in km) for $a=1.7$ and $b=0.7$.}
    \label{fig:7}
\end{figure}

One of the most crucial sources of understanding about a gravastar's stability and discovery is the study of its surface redshift. Surface redshift must not exceed 2 for the isotropic compact star fluid (surface redshift must not exceed 5 for spacetimes with the current cosmic constant). We have employed the following equation to get the surface redshift:
\begin{equation}
    Z_s=-1+\frac{1}{\sqrt{g_{tt}}}.
    \label{eq:52}
\end{equation}
Consequently, we solve the Eq. \eqref{eq:52} numerically as usual and display the result in Fig. \eqref{fig:7} for PSR J1416-2230. As can be seen, surface redshift is positive for positive values of $a$ and $b$ and $z_s<1$.  Therefore, it can be said that our current research on gravastar is both steady and physically acceptable.

\section{Junction condition and equation of state}\label{sec:5}

\begin{figure}[h]
\advance\leftskip-1.3cm
\includegraphics[scale=0.456]{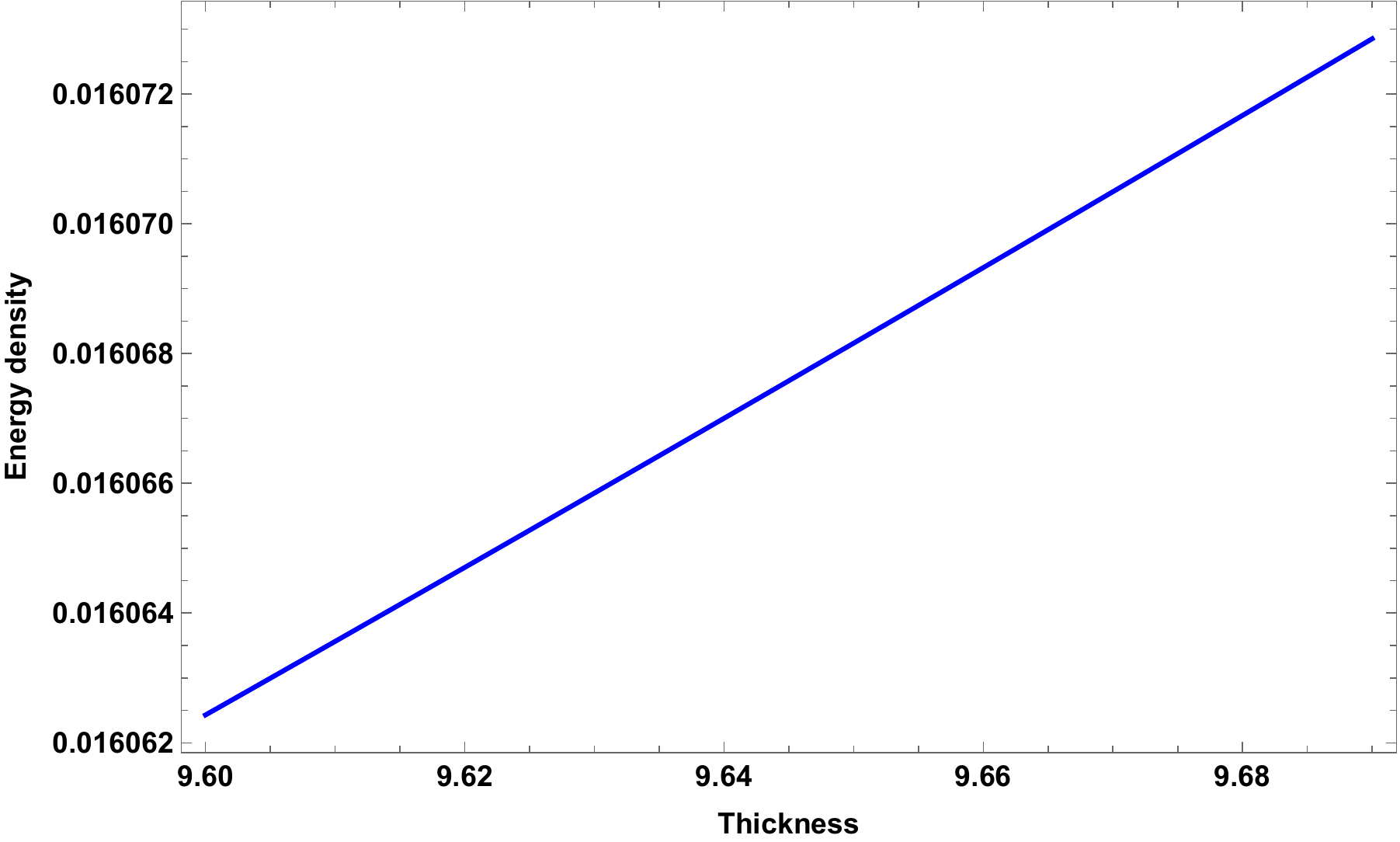}
    \caption{Graphical analysis of the surface energy density of the shell with its thickness($\epsilon$ in km) for $a=1.7$ and $b=0.7$.}
    \label{fig:8}
\end{figure}

The three regions that gravastars are known to possess are the interior region, exterior area, and shell. This shell serves as a junction between the inside and exterior space. Consequently, it is highly important to the gravastar arrangement.  According to the basic junction condition, there must be a seamless matching between regions I and III at the junction. According to the work of Darmois-Israel, \cite{darmois, israel} metric coefficients are continuous at the junction surface, it's possible that the derivatives of these metric coefficients are not. Thus, the following definition can be used to express the stress-energy surface tensor at the junction:

\begin{align}
    S_{\alpha \beta}=-\frac{1}{8 \pi}(k_{\alpha \beta}-\delta_{\alpha \beta} k_{\gamma \gamma}),
\end{align}
where $k_{\alpha \beta}=K^{+}_{\alpha \beta}-K^{-}_{\alpha \beta}$. The second elemental form is provided by a thin shell with two sides given by,
\begin{equation}
    K^{\pm}_{ij}=-n^{\pm}_{\sigma} \left(  \frac{\partial x_{\sigma}}{\partial \phi^{\alpha} \partial \phi^{\beta}} +\Gamma^{i}_{kj} \frac{\partial x^i}{\partial \phi^{\alpha}} \frac{\partial x^j}{\partial \phi^{\beta}} \right),
\end{equation}
where $\phi^{\alpha}$ denotes the intrinsic co-ordinate on the shell and for two-sided unit normals to surface, $n^{\pm}$ is represented as what can be written as,
\begin{equation}
    n^{\pm}=\pm \left|g^{ij} \frac{\partial f}{\partial x^{i}} \frac{\partial f}{\partial x^{j}} \right|^{-1/2} \frac{\partial f}{\partial x^{\sigma}},
\end{equation}
with $n^{\gamma} n_{\gamma} =1$.
 By applying the Lanczos method \cite{lanczos} the surface energy tensor can be found as $S_{\alpha \beta}=diag(-\sum, P)$, where the surface energy density and surface pressure are denoted by  $\sum$   and $P$ respectively and are defined by,

 \begin{equation}
     \sum=-\frac{1}{4 \pi R} \left[\sqrt f\right]^{+}_{-},
     \label{eq:55}
 \end{equation}
 \begin{equation}
     P=-\frac{\sum}{2}+\frac{1}{16 \pi}\left[\frac{f'}{\sqrt f}\right]^{+}_{-}.
     \label{eq:56}
 \end{equation}
Where $f$ is the metric potential term $e^{\lambda}$. Equation \eqref{eq:55}, \eqref{eq:56} turns into the following form in our calculation as,
 \begin{equation}
     \sum=-\frac{1}{4 \pi R} \left[ 
  \sqrt{1-\frac{2M}{R}}-\sqrt{1+\frac{R^2}{6a }( b-16 \pi \rho_c)}\right]
  \label{eq:58}
 \end{equation}

 \begin{multline}
    P=\frac{1}{16 \pi }\left(\frac{2 M}{\sqrt{1-\frac{2 M}{R}} R^2}-\frac{R (b  -16  \pi{\rho_c})}{3 a  \sqrt{1+\frac{R^2 (b
 -16 {\pi \rho_c})}{6 a }}}\right)\\
-\frac{1}{2} \left(-\frac{1}{4 \pi  R}\right)\left(\sqrt{1-\frac{2 M}{R}}-\sqrt{1+\frac{R^2 (b  -16
{\pi \rho_c})}{6 a  }}\right)
\label{eq:59}
 \end{multline}
 Also, the EoS parameter $\omega$ takes the form,\\
 \begin{equation}
     \omega=\frac{P}{\sum}.
 \end{equation}
 Which implies,\\
 \begin{equation}
     \omega=-\frac{1}{2}-\frac{R \left(\frac{2 M}{\sqrt{1-\frac{2 M}{R}} R^2}-\frac{R (b  -16 \pi\rho_c)}{3 a   \sqrt{1+\frac{R^2
(b  -16 \pi \rho_c)}{6 a  }}}\right)}{4 \left(\sqrt{1-\frac{2 M}{R}}-\sqrt{1+\frac{R^2 (b -16  \pi \rho_c)}{6 a }}\right)}.
\label{eq:61}
 \end{equation}
 It is noted that for the real value of Eos parameter $\omega$, the quantity $\frac{2M}{R}<1$ also $\frac{R^2(b  -16 \pi \rho_c)}{6a} \neq -1$ same to our previous condition regarding the solution in Eq. \eqref{eq:33} for the interior region.
 The graphical analysis of the surface energy density  has given in Fig. \ref{fig:8}. It shows positive behavior and increases monotonically with the increases of thickness $\epsilon$.
 
 At last, we have calculated the mass of the thin shell ($m_s$) by applying the formula,
\begin{multline}
    m_s=4 \pi R^2 \sum \\
    =-R\left[ 
  \sqrt{1-\frac{2M}{R}}-\sqrt{1+\frac{R^2}{6a}( b-16 \pi \rho_c)}\right].
  \label{eq:62}
\end{multline}

Thus the total gravitational mass ($M$) of the gravastar can be determined in terms of shell mass($m_s$) as,

\begin{multline}
   M= \frac{1}{12 a  R} \bigg[  -6 a {m_s}^2  -b   R^4+16 \pi R^4 {\rho_c}\\+2 a {m_s} \sqrt{6  } R \sqrt{\frac{6 a  +b  R^2-16  \pi R^2
{\rho_c}}{a}} \bigg].
\label{eq:63}
\end{multline}

\section{discussion and conclusion on the status of the Gravastar model}\label{sec:6}

There are several theories of gravity in which the gravastar has been studied. But, in this work, we are attempted to study gravastar in a novel gravitational framework called symmetric teleparallel gravity. The motivation for working in this gravitational framework is that it is very successful in describing the current accelerated expansion of the universe, and also it is tested by the solar system test to be a self-consistent theory. To proceed further, we have presumed the static and spherically symmetric metric with the Kuchowicz metric potential to present physically feasible, singularity-free (to avoid the black hole singularity-like nature of the compact spherically symmetric astronomical objects) gravastars. Using this metric with an ideal fluid distribution, we have examined different features of the interior, shell, and exterior regions. Further, we have analyzed various physical properties of those regions precisely in the framework of $f(Q)$ gravity as follows:
\begin{itemize}

 \item  \textbf{Interior region:} Using the EOS for the interior region, motion equations, and conservation equation, we have obtained that the solution is free from the singularity, and the energy density and pressure remain constant with dark energy nature. 

 \item \textbf{Intermediate thin shell:} We have used the intermediate shell condition for the matter profile and found the metric potential for it. Profile of the obtained metric potential is presented graphically in Fig. (\ref{fig:2}). Gravastar shell plays a major role in describing their structure and differentiating it from black hole. Therefore, we have studied various physical properties. The details are presented in the following:
  \end{itemize}
  
\begin{enumerate}

 \item \textbf{Pressure and matter density:} For the shell region, we have found the solution for matter density and pressure. The matter density profile with respect to the thickness of the shell have presented in Fig.(\ref{fig:3}), and it is shown that it decreases with an increase in thickness.

 \item \textbf{Proper length:} The numerical profile of the proper length with the change of thickness parameter $\epsilon$ has depicted in Fig(\ref{fig:4}). It is observed that when thickness values increase, the proper length of the shell steadily decreases.

 \item \textbf{Energy:} From Fig.(\ref{fig:5}), it is discovered that the energy variation for the shell is positive and monotonically increases toward the outer surface. This implies that the shell's outer boundary is denser than its internal limit. It meets with the condition that the shell's energy grows as the radial distance increases.

 \item \textbf{Entropy:} The evolution of the shell entropy has shown in Fig.(\ref{fig:6}). It is noted that the entropy of the shell decreases as its thickness increases in $f(Q)$ gravity. This behavior, however, does not suggest that our solution is unstable.

 \item \textbf{Surface Redshift:} For PSR J1416-2230, we have computed the numerical results for surface redshift and displayed the result in Fig.(\ref{fig:7}). As it can be seen, surface redshift is positive for positive values of $a$ and $b$ and $Z_s<1$. Therefore, it can be said that our current research on gravastar is both steady and physically acceptable.

 \item \textbf{Exterior region:} There are two limits in the gravastar configuration: one is between the interior and intermediate thin shell at a distance $R_1$, and the other is between the shell and outer space at a distance $R_2$ from the center. The metric functions at these interfaces must be continuous for any stable system. Therefore, we have matched these metric potentials with outer space-time.

 \item \textbf{Junction Condition and EoS:}
 According to the basic junction condition, there must be a seamless matching between regions I and III at the junction. According to the work of Darmois-Israel, \cite{darmois,israel} metric coefficients are continuous at the junction surface. We have derived the formula of energy density \eqref{eq:58} and pressure \eqref{eq:59} at the surface. Along with that, we have shown the physical behavior of energy density in Fig. \ref{fig:8}, which describes the increasing behavior of surface energy density with respect to the thickness of the shell. We have restricted some conditions for having real values of EoS parameter $\omega$ like  $\frac{2M}{R}<1$ also $\frac{R^2(b  -16 \pi\rho_c)}{6a } \neq -1$ same to our previous condition regarding the solution in Eq. \eqref{eq:33} for the interior region. Finally, we have derived the expression of the mass of the thin shell ($m_s$) as well as the total mass ($M$) in Eq. \eqref{eq:62} and Eq. \eqref{eq:63}, respectively.
 
\end{enumerate}

Gravastar is a theoretically motivated spherically symmetric compact astrophysical object to solve the singularity issue in the black hole geometry. Now, the question arises about the physical existence and discovery of the gravastar in our universe. Despite the fact that grvastars have not yet been observed or found scientifically, However, there are a number of reasons for and against the theory that gravitational waves (GW) detected by LIGO are the result of merging gravastars or black holes. The theoretical existence of Gravastar and all of its physical feasibility are demonstrated in this article. 

Finally, we can conclude that in the present paper, a successful study was done on gravastar in the context of the symmetric teleparallel theory of gravity with the help of Kuchowickz metric potential. We obtained a set of physically acceptable and non-singular solutions of the gravastar, which immediately overcame the problem of the central singularity and the existence of the event horizon of the black hole. Analyzing all the results we obtained, we claim the possible existence of gravastar in symmetric teleparallel theory as obtained in Einstein's GR.

\vspace{1 cm}
\textbf{Data availability:} There are no new data associated with this article.

\section*{Acknowledgement}
SP \& PKS  acknowledges the National Board for Higher Mathematics (NBHM) under the Department of Atomic Energy (DAE), Govt. of India for financial support to carry out the Research project No.: 02011/3/2022 NBHM(R.P.)/R \& D II/2152 Dt.14.02.2022.

\end{CJK*}
\end{document}